\documentclass[aps,article,nofootinbib,groupedaddress,twocolumn]{revtex4-1}
\usepackage{graphicx}
\usepackage{amsmath,amssymb}
\usepackage{hyperref}
\usepackage[usenames]{color}
\usepackage{url}
\usepackage{epstopdf}

\hypersetup{
    colorlinks=true,
    linkcolor=black,
    citecolor=blue,
    urlcolor=black
}

\def\be{\begin{equation}}
\def\ee{\end{equation}}
\def\ba{\begin{eqnarray}}
\def\ea{\end{eqnarray}}

\def\f{\frac}
\def\l{\left}
\def\r{\right}
\def\hub{{\cal H}}

\begin{document}

\title{Practical solutions for perturbed $f(R)$ gravity}
\date{\today}

\author{Alireza Hojjati$^{1,2}$, Levon Pogosian$^{1,3}$, Alessandra Silvestri$^{4,5}$, and Starla Talbot$^{1}$}

\affiliation{$^1$Department of Physics, Simon Fraser University, Burnaby, BC, V5A 1S6, Canada \\
$^2$ Institute for the Early Universe, Ewha Womans University, Seoul, 120-750, South Korea\\
$^3$Centre for Theoretical Cosmology, DAMTP, University of Cambridge, CB3 0WA, UK\\
$^4$ Department of Physics, MIT, Cambridge, MA 02139, USA \\
$^5$ SISSA - International School for Advanced Studies, Via Bonomea 265, 34136, Trieste, Italy}

\begin{abstract}
We examine the accuracy of the quasi-static approximation and a parametric form commonly employed for evolving linear perturbations in $f(R)$ gravity theories and placing cosmological constraints. In particular, we analyze the nature and the importance of the near horizon effects that are often neglected. We find that for viable models, with a small present value of the scalaron field, such corrections are entirely negligible with no impact on observables. We also find that the one-parameter form, commonly used to represent the modified equations for linear perturbations in $f(R)$, leads to theoretical systematic errors that are relatively small and, therefore, is adequate for placing constraint on $f(R)$ models.
\end{abstract}

\maketitle

\section{Introduction}

Understanding the physics behind the observed accelerating expansion of the 
universe~\cite{Riess:1998cb,Perlmutter:1998np} is one of the most important task of modern science. The simplest theoretical explanation within the context of General Relativity (GR)~\cite{einstein1915} is the cosmological constant $\Lambda$, which also happens to be in good agreement with the current cosmic microwave background (CMB)~\cite{Komatsu:2010fb}, supenovae (SN)~\cite{Conley:2011ku,Suzuki:2011hu}, and baryon acoustic oscillations (BAO)~\cite{Blake:2011en,Reid:2012sw} data. However, the observed $\Lambda$ is many orders of magnitude smaller than the vacuum energy density predicted by quantum field theory~\cite{Weinberg:1988cp}, implying either an extraordinary degree of fine-tuning or an anthropic argument~\cite{Weinberg:1987dv,Vilenkin:1994ua}. Many alternatives to $\Lambda$ have been proposed in the literature so far, and they can be broadly divided into two classes: dynamical dark energy~\cite{Copeland:2006wr} and modified gravity (MG)~\cite{Silvestri:2009hh,Clifton:2011jh}. 

A very popular model of MG involves adding a function of the Ricci scalar, $f(R)$, to the Einstein-Hilbert action of GR, such that the resulting equations for the metric admit self-accelerating solutions~\cite{Capozziello:2003tk,Carroll:2003wy,Hu:2007nk,Appleby:2007vb,Starobinsky:2007hu}. In fact, given the $4$th order nature of the $f(R)$ equations of motion, it is possible to design $f(R)$ functions to match any expansion history of the universe~\cite{Song:2006ej}. Prior to its appearance in the context of the current cosmic acceleration, $f(R)$ was introduced by Starobinsky as the first working model of Inflation~\cite{Starobinsky:1979ty,Starobinsky:1980te}.

The $4$th order nature of the equations implies the existence of an additional scalar degree of freedom coupled to all matter with gravitational strength\footnote{By transforming to a conformally related frame $f(R)$ can be explicitly written as a scalar-tensor gravity theory~\cite{Magnano:1993bd,Chiba:2003ir}.}. 
The scalar field, dubbed ``scalaron'', mediates an attractive force, with the trace of the energy-momentum acting as charge. It can evade solar system tests~\cite{CMWill} constraining fifth force interactions thanks to the Chameleon mechanism~\cite{Khoury:2003aq} in which the scalaron is screened in high density environments. While most choices of the $f(R)$ function result in a time-varying equation of state $w$ for the effective dark energy fluid, viable models are forced to have an expansion history that is practically indistinguishable from that of the $\Lambda$ Cold Dark Matter (LCDM) model with $w=-1$~\cite{Dolgov:2003px,Amendola:2006kh,Nojiri:2006gh,Hu:2007nk,Appleby:2007vb,Starobinsky:2007hu}. There can, however, be detectable differences between $f(R)$ and LCDM when it comes to the dynamics of clustering of matter, both on linear and non-linear scales~\cite{Song:2006ej,Hu:2007nk,Pogosian:2007sw,Oyaizu:2008tb,Schmidt:2008tn,Zhao:2011cu,Ma:2011hf,Li:2011pj,Lombriser:2012nn,Li:2012by}. Searching for evidence of modified growth dynamics is indeed one of the main science goals of upcoming and future large scale surveys such as DES~\cite{DES}, Euclid~\cite{Euclid} and LSST~\cite{LSST}. 

The dynamics of linear cosmological perturbations in $f(R)$ has been studied in~\cite{Bean:2006up,Song:2006ej,Pogosian:2007sw}. In principle, the non-linearity of the $f(R)$ function means that it is not guaranteed that thepredictions of perturbation theory will necessarily agree with the actual dynamics of clustering on large scales. In fact, if all of the matter was in sufficiently dense form, the scalaron force would always be suppressed by the Chameleon mechanism and the growth on linear scales would be identical to that in GR. In reality, most of the matter in the universe resides in diffuse regions that do not screen the fifth force, and N-body simulations~\cite{Oyaizu:2008sr,Oyaizu:2008tb,Schmidt:2008tn,Zhao:2010qy,Li:2011pj} have confirmed that predictions of linear theory are indeed recovered on large scales. Given the degree of fine-tuning required to design viable $f(R)$ models, they can hardly be considered a serious alternative to $\Lambda$. However, they provide a simple working toy-model for studying strongly coupled scalar-tensor theories with Chameleon screening that help us better understand ways in which gravity can be tested with upcoming observational data.

Solving exact perturbed equations  in MG models for the purpose of calculating observables can be challenging. Often authors rely on approximate parametrized forms of the equations which are meant to capture the main physical features of the model while providing a simplified framework to evolve perturbations. A fairly general five-parameter model was introduced by Bertschinger and Zukin (BZ) in~\cite{Bertschinger:2008zb}. A single parameter version of the BZ parametrization was used in~\cite{Zhao:2008bn,Giannantonio:2009gi,Hojjati:2011ix} and other literature to mimic the solutions of the perturbation equations in $f(R)$ theories when deriving cosmological constraints. However, the accuracy of this single parameter form has not been consistently tested. In this work, we compare the single-parameter form used in~\cite{Zhao:2008bn,Giannantonio:2009gi,Hojjati:2011ix} to the exact numerical solutions in $f(R)$ and estimate the systematic theoretical error. We analyze several aspects of this parametrization, including the accuracy of the quasi-static approximation on which it is based. We find that the near-horizon effects are more prominent for large values of the scalaron today, which are already ruled, but are negligible for $f(R)$ models in the viable range. Overall, we conclude that the single parameter BZ form is an adequate representation of solutions in linearized $f(R)$ models for the purpose of testing them with the most advanced future surveys. 

\section{$f(R)$ theories}
\label{Sec:II}

We consider $f(R)$ models of gravity in the Jordan frame, for which the action reads:
\be
S=\frac{1}{2M_P^2}\int d^4 x\sqrt{-g}\, \left[R+f(R)\right] +S_{\rm m}[\chi_i,g_{\mu\nu}] \ ,
\label{jordanaction}
\ee
where $f(R)$ is a general function of the Ricci scalar. Matter is minimally coupled and, therefore, the matter fields, $\chi_i$, fall along geodesics of the metric $g_{\mu\nu}$. The field equations obtained from varying action~(\ref{jordanaction}) with respect to $g_{\mu\nu}$ are of higher order than those in GR. At the background level, the cosmological evolution is described by the Friedmann equation:
\be\label{jordanfriedmann}
(1+f_{R})\hub^2+\frac{a^{2}}{6}f-\f{\ddot{a}}{a}f_{R} +\hub \dot{f}_{R} =  \f{a^{2}}{3M_P^2}\rho\,,
\ee
and the trace equation:
\be\label{trace}
\Box f_R=\f{1}{3}\l(R+2f-Rf_R\r)-\f{\rho-3P}{3M_P^2} \, ,
\ee
where a dot indicates derivation w.r.t. conformal time, $\hub$ is the Hubble parameter in conformal time, and we have defined $f_R\equiv \partial f/\partial R$.

Eq.~(\ref{trace}) can be interpreted as an equation of motion for a scalar field $f_R$, dubbed {\it scalaron} in~\cite{Starobinsky:1979ty}, with an effective potential that depends on the density of matter and gives the scalaron an effective mass\be\label{mass_scalaron}
m^2_{f_R}=\f{1}{3}\l[\f{1+f_R}{f_{RR}}-R\r]\approx\f{1+f_R}{3f_{RR}} \ .
\ee
There are certain conditions that need to be imposed on $f(R)$ theories in order to avoid instabilities and reproduce the correct high curvature regime. Namely, one needs $f_{RR}>0$ and $f_{RR} R \ll 1$ to have a stable high curvature regime and a non-tachyonic scalaron~\cite{Dolgov:2003px}, and $1+f_R>0$ to prevent the scalaron from turning into a ghost~\cite{Nariai:1973eg,Gurovich:1979xg,Nunez:2004ji}. Considering that we want to reproduce GR at early times, i.e. we want $f_R\rightarrow 0$ as $R\rightarrow \infty$, these conditions imply that $f_R$ will be a monotonically increasing function of $R$ that asymptotes to zero from below. If one furthermore wants to satisfy existing constraints from local tests of gravity, $|f_R|$ has to be small at recent epochs; solar system and nearby universe tests impose a bound of $|f_R|\lesssim 10^{-6}$~\cite{Hu:2007nk,Jain:2012tn}.

The Friedmann equation~(\ref{jordanfriedmann}) can be re-written as a second order equation for $f(a)$. Given an expansion history $H(a)$, and hence $R(a)$, one can use (\ref{jordanfriedmann}) to solve for $f(a)$ and then determine $f(R)$~\cite{Song:2006ej,Pogosian:2007sw}. In this ``designer'' approach, the amplitude of the decaying mode is set to zero, while the cosmologically relevant growing mode solution is not unique --  for each expansion history there is a family of $f(R)$ models parametrized by a boundary condition, such as the value of $\partial f/\partial R$ today, which we denote as $f_R^0$. In the literature, constraints on $f(R)$ are often presented in terms of a parameter $B_0$~\cite{Song:2006ej}, which is related to the mass of the scalaron today. It is another way of setting the boundary condition that labels the models and is given by the value of the function 
\be\label{B}
B\equiv\f{f_{RR}}{1+f_R}\f{\hub \dot{R}}{\dot{\hub}-\hub^2}
\ee
today. In a given $f(R)$ model, $B_0$ is in a one-to-one correspondence with $f_R^0$ -- a heavier scalaron (smaller $B_0$) corresponds to a smaller value of $|f_R^0|$. In models with a LCDM expansion history and $\Omega_M=0.27$, they are approximately related as $B_0\approx -6 f_R^0$ over the range of interesting values of $f_R^0$. Generally, models satisfying the viability conditions and the local tests of gravity discussed above, predict an expansion history practically indistinguishable from LCDM~\cite{Dolgov:2003px,Amendola:2006kh,Nojiri:2006gh,Hu:2007nk,Appleby:2007vb,Starobinsky:2007hu}. Still, the dynamics of perturbations can differ significantly~\cite{Song:2006ej,Pogosian:2007sw}.

To describe the growth of structure and evolution of the CMB anisotropies, we expand the $f(R)$ field equations to first order in perturbations. The linearly perturbed Einstein equations contain terms in $\delta f_R$ and its first and second time derivatives, making them fourth order in derivatives of the metric (as opposed to the second order of GR equations). We will not reproduce the full set of exact equations here, as only the {\it Poisson} and the {\it anisotropy} equations will be relevant for our discussion. In what follows, we will compare the latter to the equivalent equations in GR to highlight the modifications introduced by $f(R)$ theories. We consider scalar perturbations in the conformal Newtonian gauge, with $\delta g_{00}=-2a^2\Psi$ and $\delta g_{ij}=-2a^2\Phi\delta_{ij}$,  and  present all our equations in Fourier space.

Let us start with the anisotropy equation, which is the spatial off-diagonal component of the linearized Einstein equation. In GR, it reads
\be
\label{an_GR}
k^2\l(\Phi-\Psi\r)=\f{3a^2}{2M_P^2}\l(\rho+P\r)\sigma \,,
\ee
where $\sigma$ is the anisotropic stress of matter. In $f(R)$ it becomes
\be
\label{an_f(R)}
k^2\l(\Phi-\Psi\r)-k^2\f{\delta f_R}{F}=\f{3a^2}{2M_P^2}\f{\l(\rho+P\r)}{F}\sigma \,,
\ee
where $F\equiv1+f_{R}$.

The Poisson equation is obtained with a suitable combination of the time-time and time-space components of the Einstein equations that leaves the comoving density contrast as the contribution from matter and, in GR, the Laplacian acting on the curvature potential as the only geometric term. It is, however, common to work with the analogous equation for the Newtonian potential, since the latter is more directly related to observables given that it governs the motion of non-relativistic particles. In GR,  combining the time-time, time-space and anisotropy equations (neglecting the anisotropic stress), one obtains the following algebraic relation: 
 \be\label{Poisson_Psi_GR}
k^2\Psi=-\f{a^2}{2M_P^2}\rho\Delta \ ,
\ee
where $\Delta\equiv \delta +3\hub/k v$ is the comoving density contrast. The latter is valid on all scales, as long as the anisotropic stress is negligible, and is commonly employed to study the clustering of matter. This is the equation that is often parametrized along with the anisotropy equation. In $f(R)$, due to the higher order nature of the theory, the same combination of Einstein equations leads to a 'modified Poisson' equation for $\Psi$, which is now dynamical and reads:
\ba
\label{Poisson_Psi_f(R)}
k^2\Psi&-&k^2\f{\delta f_R}{2F}+\f{3}{2}\l[\l(\dot{\hub}-\hub^2\r)\f{\delta f_R}{F}+\l(\dot{\Phi}+\hub\Psi\r)\f{\dot{F}}{F}\r]\nonumber\\
&&=-\f{a^2}{2M_P^2}\f{\rho}{F}\Delta \,.
\ea
Using $\delta f_R=f_{RR}\delta R$, expanding $\delta R$ in terms of the metric potentials, and applying the quasi-static approximation, Eq.~(\ref{Poisson_Psi_f(R)}) reduces to a simple generalization of~(\ref{Poisson_Psi_GR}), which will be discussed in Sec.~\ref{Sec:IIA}.

It is often preferable to work with a Poisson equation for the lensing potential, governing the motion of relativistic particles.  In GR, when neglecting anisotropic stress, the latter reads:
\be\label{Poisson_Weyl_GR}
k^2\l(\Phi+\Psi\r)=-\f{a^2}{M_P^2}\rho\Delta\,.
\ee
This equation is again valid on all scales and is convenient to use when one wants to study weak lensing and the Integrated Sachs-Wolfe (ISW) effect. The equivalent equation in $f(R)$ is again dynamical:
\ba\label{Poisson_Weyl_f(R)}
k^2(\Phi+\Psi)&+&3\l[\l(\dot{\hub}-\hub^2\r)\f{\delta f_R}{F}+\l(\dot{\Phi}+\hub\Psi\r)\f{\dot{F}}{F}\r]\nonumber\\
&&=-\f{a^2}{M_P^2}\f{\rho}{F}\Delta \,.
\ea
In the quasi-static sub-horizon limit, when the terms in the square bracket become negligible, Eq.~(\ref{Poisson_Weyl_f(R)}) reduces to a trivial generalization of~(\ref{Poisson_Weyl_GR}), {\it i.e.} a rescaling of the Newton constant by $F$.

Overall, the fourth order nature of the linearly perturbed equations in $f(R)$ theories implies a richer dynamics and potential instabilities, just like at the background level.  In particular, at early times, {\it i.e.} high curvature and high scalaron mass, the equation for $\delta f_R$ has highly oscillatory solutions~\cite{Starobinsky:2007hu,Song:2006ej,Pogosian:2007sw} with a frequency proportional to the scalaron mass. As long as $f_{RR}>0$ (assuming $F>0$), these oscillations remain small in amplitude and decay in time (see Sec.~\ref{Sec:IIA}) ensuring that instabilities do not form in the evolution of linear perturbations. 

\subsection{Linear growth in $f(R)$ and its approximate parametrizations}\label{Sec:IIA}

As discussed above, Einstein's equations in $f(R)$  theories are quite different from those in GR and can result in significantly different dynamics of linear perturbations. But modifications of linear growth can also occur in models of exotic dark energy and dark matter that are otherwise based on GR. Hence, we use ``MG'' to denote not only modified gravity, but also, more generally, {\it modified growth}.

An arbitrary modification to the dynamics of scalar perturbations on linear scales can be encoded into two time- and scale-dependent functions, $\mu(a,k)$ and $\gamma(a,k)$,  generalizing the Poisson~(\ref{Poisson_Psi_GR}) and anisotropy~(\ref{an_GR}) equations of GR to the following:
\ba\label{Poisson_Modified}
&& k^{2}\Psi = -\frac{a^{2}}{2M_{p}^{2}}\mu(a,k)\rho\Delta \,,\\  
\label{an_Modified}
&&\frac{\Phi}{\Psi} = \gamma(a,k)\,,
\ea
where any deviation of $\mu$ and $\gamma$ from unity signals a departure from the LCDM growth at late times\footnote{Several other choices of the two functions parametrizing perturbed MG equations, equivalent to $\mu$ and $\gamma$, can be found in the literature. For a summary see~\cite{Daniel:2010ky,Pogosian:2010tj}.}. The functions $\mu$ and $\gamma$ do not necessarily have a simple form in specific models of MG. Strictly speaking, they parametrize solutions of the equations of motions and depend on the choice of the initial conditions\footnote{This is true also for the conceptually similar but technically different approach proposed in~\cite{Hu:2007pj}. Alternative parametrization schemes that are more directly related to particular theories have also been developed, {\it e.~g.} in~\cite{Baker:2012zs,Battye:2012eu,Brax:2012gr}. They are either significantly more complex, as in \cite{Baker:2012zs,Battye:2012eu}, or apply to a more limited range of models~\cite{Brax:2012gr}.}. Nevertheless, they provide a consistent framework for searching for departures from LCDM. In some theories, including $f(R)$, they can assume a simple form on sub-horizon scales. However, one can also study them in a model-independent way. For instance, in~\cite{Zhao:2009fn,Hojjati:2011xd,Hojjati:2012ci}, a principal component analysis of these function was used to forecast and analyze the ability of experiments like DES and LSST to constrain MG. 

Depending on the observables that one considers, it is sometimes convenient to parametrize the Poisson equation for the lensing potential, rather than the one for $\Psi$~\cite{Hojjati:2011xd,Zhao:2010dz,Song:2010fg}. Namely, one can introduce a function $\Sigma(a,k)$ via
\be\label{Poisson_Weyl_Modified}
k^2\l(\Phi+\Psi\r)=-\f{a^2}{M_P^2} \Sigma(a,k) \rho\Delta\ ,
\ee
which is more directly probed by statistical quantities derived from weak lensing of distant galaxies.

In $f(R)$, the Compton wavelength associated with the mass of the scalaron (\ref{mass_scalaron}) sets the range of the fifth force interaction, which separates two regimes of dynamics. On scales larger than the Compton length, modifications are negligible and the dynamics is very close to that in GR. Below the Compton scale the growth is enhanced and the two metric potentials are no longer equal. This scale-dependent behavior can be easily seen from the Poisson~(\ref{Poisson_Psi_f(R)})  and anisotropy~(\ref{an_f(R)}) equations in the quasi-static sub-horizon limit, as discussed at length in~\cite{Pogosian:2007sw}. Assuming that on sub-horizon scales ($k\gg aH$), the time variation of the gravitational potentials is slow compared to their variation in space, one obtains a simplified form of the equations that can be reproduced substituting the following functions into~(\ref{Poisson_Modified}) and~(\ref{an_Modified}):
\ba
\label{muQ}
&& \mu^{Q}(a,k)=\frac{1}{F}\frac{1+(4/3)Q}{1+Q} \, \\
\label{gamQ}
&& \gamma^{Q}(a,k) = \frac{1+(2/3)Q}{1+(4/3)Q} \,,
\ea
where the dimensionless parameter $Q$~\cite{Pogosian:2007sw} is  the squared ratio of the scalaron Compton wavelength, $\lambda_C\equiv 2\pi/m_{f_R}$ (with $m_{f_R}$ given by~(\ref{mass_scalaron})), to the physical wavelength associated with $k$:
\ba
\label{Q}
Q\equiv 3\frac{k^{2}}{a^{2}}\frac{f_{RR}}{F}\approx \left(\frac{\lambda_{C}}{\lambda}\right)^{2}\,.
\ea
Note that the $1/F$ factor in~(\ref{muQ}), which enhances the growth in a scale-independent way, is small for values of $|f_R^0| \ll 1$, favored by local tests of gravity, but is not entirely negligible for larger values. Also, for $f(R)$ in the quasi-static sub-horizon limit, $\Sigma(a,k)$ reduces to a simple scale-independent expression, {\it i.e.} 
\be
\Sigma^{Q}(a)=1/F \ .
\ee
As evident from Eq.~(\ref{Poisson_Weyl_f(R)}), there are extra terms that would modify this simple relation on larger scales.

The quasi-static expressions for $\mu$~(\ref{muQ}) and $\gamma$~(\ref{gamQ}), motivated the following general parametrization introduced by Bertschinger and Zukin (BZ) in~\cite{Bertschinger:2008zb}
\ba\label{muBZ1}
&& \mu(a,k) = \frac{1 + \alpha_1k^{2}a^{s}}{1+\alpha_{2}k^{2}a^{s}} \,, \\
\label{gamBZ1}
&& \gamma(a,k) = \frac{1+\beta_1k^{2}a^{s}}{1+\beta_2k^{2}a^{s}} \, .
\ea
It contains $5$ parameters $(\alpha_1,\alpha_2, \beta_1,\beta_2,s)$ and describes a transition of both functions from one constant value to another, analogously to what happens in the quasi static limit of $f(R)$. In fact, the correspondence with Eqs.~(\ref{muQ})-(\ref{gamQ}) would be exact if the scalaron Compton length had a power law dependence on the scale factor, $a$. We will adopt this as an approximation for now and examine its implications later. As discussed earlier, for a given expansion history, $f(R)$ theories have a single free parameter, $f_R^0$, or equivalently, $B_0$. It is easy to express analytically $\alpha_1$, $\alpha_2$, $\beta_1$ and  $\beta_2$ in terms of $B_0$ and one is then left with the time dependence, parametrized by $s$, which can be determined numerically once the $f(R)$ model has been reconstructed. 

In~\cite{Giannantonio:2009gi}, the authors suggested a slight generalization of~(\ref{muBZ1})-(\ref{gamBZ1}) to include the pre-factor $1/F$ whose effect can be important for ISW at large $B_0$. Combining these facts, we consider the following expressions, which we will refer to as the $BZ$ parametrization:
\ba\label{muBZ}
&& \mu^{\rm BZ}(a,k) ={1 \over 1- B_0 a^{s-1}/6} \left[ \frac{1 + (2/3) B_0 \bar{k}^{2}a^{s}}{1+(1/2) B_0\bar{k}^{2}a^{s}} \right] \,, \\
\label{gamBZ}
&& \gamma^{\rm BZ}(a,k) =  \frac{1+ (1/3)B_0\bar{k}^{2}a^{s}}{1+(2/3)B_0\bar{k}^{2}a^{s}} \,,
\ea
where we have introduced a dimensionless wavenumber $\bar{k}\equiv k/H_0$, such that $k=2997.9\, \bar{k}$ Mpc/h. Comparing~(\ref{muBZ})-(\ref{gamBZ}) with~(\ref{muQ})-(\ref{gamQ}), it is easy to see that we have effectively set $\lambda_C^2\approx 2 \pi^2/H_0^2\,\, B_0\, a^{s+2}$. We have also used the numerically found approximate relation $B_0 \approx  - 6 f_R^0$ in the $1/F$ pre-factor.

In principle, the time dependence of the Compton length depends on $B_0$ and, strictly speaking, $s$ is not an independent parameter.
Indeed, the validity of the power law assumption was questioned in~\cite{Thomas:2011}. On the other hand, after it was suggested in~\cite{Zhao:2008bn} that $f(R)$ models which closely mimic LCDM correspond to $s\approx 4$, many authors have used~(\ref{muBZ})-(\ref{gamBZ}) with $s=4$ to place constraints on $f(R)$ theories in terms of bounds on $B_0$.  One of the aims of this paper is to investigate the accuracy of $s = 4$ and we show in Sec.~\ref{Sec:III} that it suffices to simply fix the value of $s$ to a constant value for the range of $f(R)$ models that will be probed with upcoming surveys.

In Sec~\ref{Sec:III}, we numerically check the accuracy of the quasi-static ($\mu^Q,\gamma^Q$) parametrization for a range of $f_R^0$ ($B_0$). We also check how well the time dependence of $Q$ is described by a power law in $a$. Additionally, we estimate the theoretical uncertainty in $f_R^0$ if one were to use $\mu^{\rm BZ}$ instead of the exact solutions. Before proceeding with these tasks, we first revisit analytically the quasi-static approximation and the potential role of near-horizon effects.

\subsection{Quasi-static approximation and near-horizon effects}
\label{Sec:QS}

To obtain the functions $\mu$ and $\gamma$ as defined in~(\ref{muQ}) and (\ref{gamQ}) one applies the quasi-static approximation which amounts to neglecting time derivatives of the perturbed quantities. Such an approximation is certainly very good on scales that are well inside the horizon, for which  $k/aH\ll1$, and still in the linear regime; it is less clear whether this approximation is safe on near horizon scales. It is convenient to work with $\delta f_R$ and $\Phi_+\equiv(\Phi+\Psi)/2$ as the two geometric scalar degrees of freedom, as in~\cite{Pogosian:2007sw}. 

If we perturb the trace equation to linear order, take the super-horizon limit and assume a heavy scalaron mass, we obtain a damped inhomogeneous harmonic equation for $\delta f_R$, similar to the one discussed in~\cite{Starobinsky:2007hu}:
\be\label{chi_SH}
\ddot{\delta f_R}+2\hub\,\dot{\delta f_R}+a^2m^2_{f_R}\delta f_R=\f{a^2}{3M_P^2}\l(\f{3\delta P}{\delta\rho}-1\r)\rho\delta\,.
\ee  
As long as $f_{RR}>0$, the solutions of this equation are damped oscillations with frequency and amplitude  proportional to the mass of the scalaron. At early times the scalaron is heavy and the oscillations have very high frequency. As time evolves, the scalaron becomes lighter and the frequency decreases, while the amplitude increases. Eventually, as the scalaron mass decreases, the damping term and the other terms that were neglected in the heavy scalaron limit of Eq.~(\ref{chi_SH}) become important, and the oscillations die off.

No oscillations are present in the lensing potential. Taking the heavy scalaron limit of the equation for $\Phi_+$,  one can see that it decouples from the equation for $\delta f_R$, becoming:

\be\label{Phiplus_SH}
\dot{\Phi}_++\hub\Phi_+=\f{a^2(\rho+P)}{2M_P^2}\f{v}{kF}
\ee
and $\Phi_+$ follows closely the evolution it would have in GR, except for the $1/F$ factor rescaling the Newton's constant.

In terms of $\delta f_R$ and $\Phi_+$, the metric potentials are given by $\Phi=\Phi_++\delta f_R/2F$ and $\Psi=\Phi_+-\delta f_R/2F$. Therefore, they display an oscillatory mode on top of the non-oscillatory term $\Phi_+$. Their oscillations are out of phase, with the same frequency as $\delta f_R$ and roughly half the amplitude (since $F\approx 1$).

Oscillations are observed also in the comoving matter density contrast, on top of the underlying growing mode. They have the same frequency as those in $\delta f_R$ and a relative amplitude which is slightly bigger than that of $\delta f_R$. This can be easily understood by looking at the $f(R)$ Poisson equation~(\ref{Poisson_Psi_f(R)}). 
 In GR, we simply have $\Delta=-2k^2/(3a^2H^2) \Psi$, valid on all scales, while Eq.~(\ref{Poisson_Psi_f(R)}) contains extra dynamical terms, which are important on super-horizon scales; in particular the term $\propto \hub^2\delta f_R$ induces oscillations on top of the growing mode with an amplitude of $\approx  H^2/(H_0^2\Omega_m(a))\delta f_R$. 

\begin{figure*} [tbhp]
\centering
\includegraphics[scale=0.4,trim=0mm 0mm 0mm 0mm]{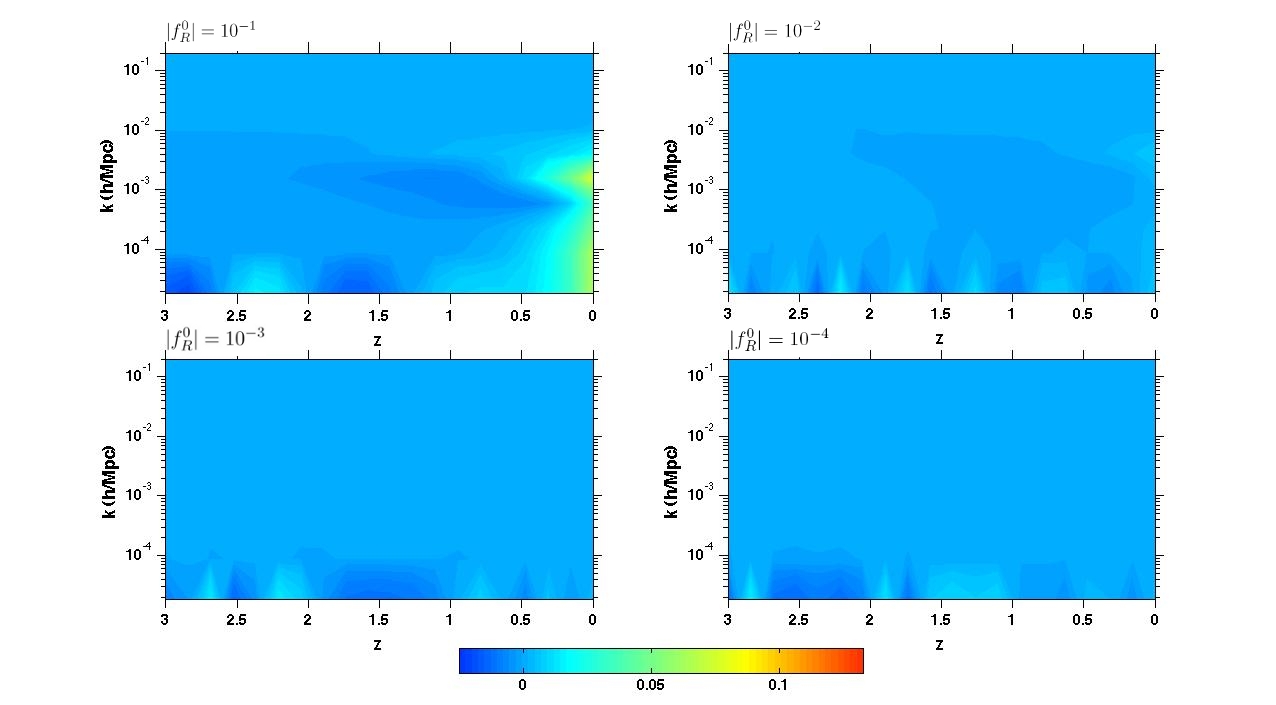}
\caption{The contour plots of the relative difference between $\mu^Q$ and $\mu^{\rm Ex}$ for four values of $f_R^0$.} 
\label{fig:muQe}
\end{figure*}

Finally, because of their definition, $\mu$ and $\gamma$, also display oscillations at low $k$ (see e.g. Fig.~\ref{fig:muQe}). While the amplitude of the oscillations in $\gamma$ increases in time (as it does for $\delta f_R, \Phi, \Psi$ and $\Delta$), the amplitude of the oscillations in $\mu$ decreases because $\Delta$ is a growing function.

These oscillatory modes, generated by dynamical terms that are important on super-horizon scales, pose certain challenges when one evolves the equations of motion numerically. On the other hand, they have practically no impact on observables. Namely, they produce effects of order $H^2/m^2_{f_R}$, which could be non negligible only for larger values of $|f_R^0|$. Moreover, they only matter on the largest scales, and the only observable that could show an imprint of these oscillations would be the ISW effect. However, the latter is sourced by the time evolution of the lensing potential which, as we saw, does not oscillate. 
Therefore, when comparing $f(R)$ theories with data, it is safe to work with the $(\mu^Q,\gamma^Q)$ parametrization obtained in the quasi-static limit, and neglect the near-horizon terms. 

\section{Parametrized vs Exact: numerical comparison}\label{Sec:III}

\begin{figure*}[tbhp]
\centering
\includegraphics[scale=0.4,trim=0mm 0mm 0mm 0mm]{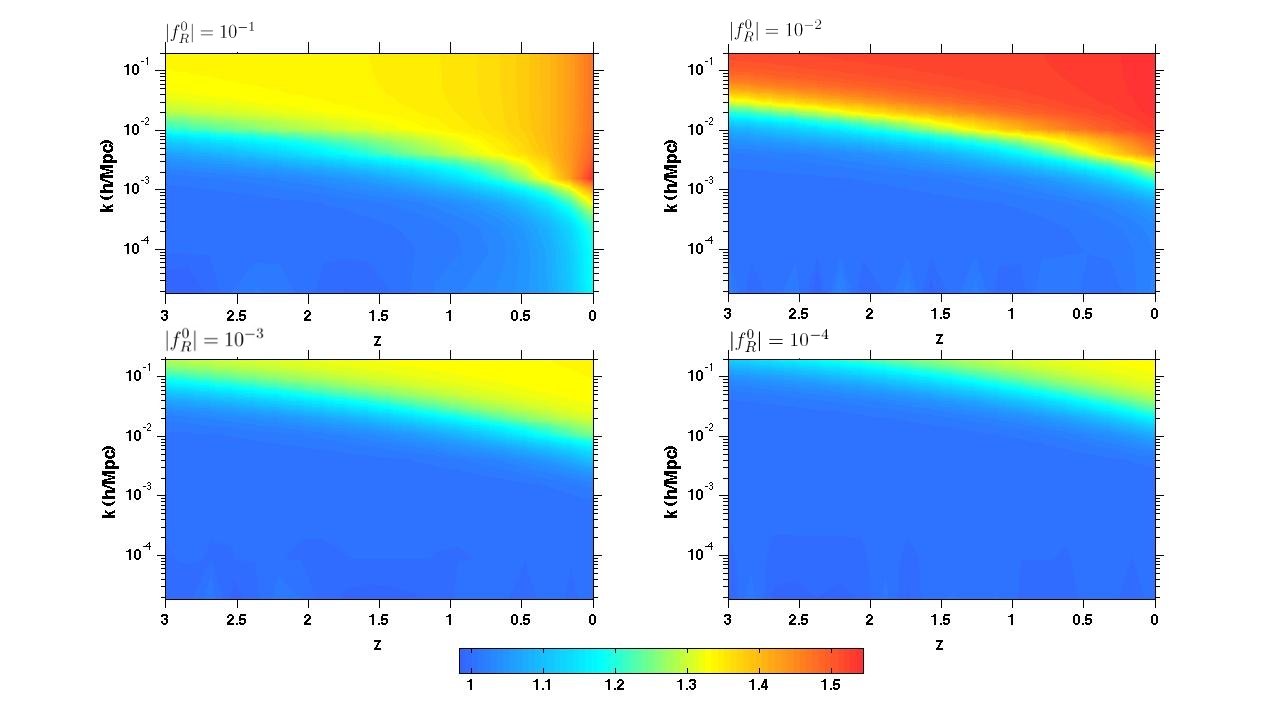}
\caption{Contour plots of $\mu^{\rm Ex}(z,k)$ obtained by numerically solving the linearly perturbed $f(R)$ equations for $|f^{0}_R|=10^{-1}$ (upper left), $10^{-2}$ (upper right), $10^{-3}$ (lower left) and $10^{-4}$ (lower right).}
\label{fig:muEx}
\end{figure*}

In this Section, we compare the functions $\mu$ and $\gamma$ obtained by solving the {\it exact} linearized $f(R)$ equations of motion to their expressions in the quasi-static approximation,~(\ref{muQ})-(\ref{gamQ}), and to the BZ representation,~(\ref{muBZ})-(\ref{gamBZ}). To obtain the ``Exact'' functions $\mu^{\rm Ex}$ and $\gamma^{\rm Ex}$, we numerically solve  the full system of linearized Einstein and energy-momentum equations, given in~\cite{Pogosian:2007sw}, to calculate $\Delta(a,k)$, $\Phi(a,k)$ and $\Psi(a,k)$. Then, we take their ratios, as prescribed by Eqs.~(\ref{Poisson_Modified}) and~(\ref{an_Modified}), to find $\mu$ and $\gamma$. 

Fig.~\ref{fig:muEx} shows contour plots of $\mu^{\rm Ex}$ in the $(z,k)$ plane, for several values of the parameter $f^{0}_R$. We opt to present the time dependence in terms of the redshift $z$, and within in the $(z,k)$ domain roughly corresponding to the range of linear scales that can be probed by future surveys. The main features of $\mu^{\rm Ex}$ are well-described by Eq.~(\ref{muQ}). On scales larger than the scalaron Compton wavelength, when $Q \ll 1$, we have $\mu=1$ because the growth of perturbations is the same as in GR. On smaller scales, $Q \gg 1$, the growth is enhanced by the fifth force and $\mu=4/3$. In addition to this scale-dependent transition, the strength of the gravitational interaction is enhanced by an overall scale-independent factor $1/F=(1+f_R)^{-1}$, which is very close to unity except for larger values of $|f^{0}_R|$. For example, the $1/F$ enhancement is clearly visible in the case of $f^{0}_R=-10^{-1}$.  

The behavior of $\gamma^{\rm Ex}$ is captured by Eq.~(\ref{gamQ}) and is similar to that of $\mu^{\rm Ex}$. We do not show plots of $\gamma^{\rm Ex}(z,k)$, noting instead that its main difference from $\mu$ is the absence of the $1/F$ factor, while the transition is from $\gamma=1$ on large scales to $\gamma=1/2$ below the Compton wavelength scale.
 
\subsection{$\mu^Q$ and $\gamma^Q$} 
\label{ssQ}

In the quasi-static regime of linear sub-horizon perturbations, the $f(R)$ modifications to the Poisson and anisotropy equations take the form of $\mu^Q$~(\ref{muQ})  and $\gamma^Q$~(\ref{gamQ}). The background functions $f_R$ and $f_{RR}$ appearing in the definition of $Q$ can be found numerically by solving the $f(R)$ Friedmann equation~(\ref{jordanfriedmann}) for a given expansion history.

To investigate the accuracy of the quasi-static approximation, we estimate the relative difference between $\mu^Q$ and $\mu^{\rm Ex}$ for values of $|f^{0}_R|$ from $10^{-1}$ to $10^{-6}$ and show the results for some of these models in
Fig.~\ref{fig:muQe}. We find that the differences decrease as $|f^{0}_R|$ gets smaller. For $|f^{0}_R| \leq 10^{-2}$ the average difference is about $0.01\%$, while the maximum difference is about $2.0\%$. The largest errors occur for $f^{0}_R = -10^{-1}$, where the maximum difference is about $7\%$. 

As the plots in Fig.~\ref{fig:muQe} show, the main difference is due the oscillations in time present on large scales in $\mu^{\rm Ex}$ and that are not present in $\mu^Q$. As discussed in Sec~\ref{Sec:QS}, $\delta f_R$ has an oscillatory behavior that translates into oscillations in $\Delta$, $\Phi$ and $\Psi$, but not in $\Phi+\Psi$. The oscillations are most prominent in the homogenous ($k\rightarrow 0$) limit and die off on smaller scales. While they dominate the relative difference between $\mu^Q$ and $\mu^{\rm Ex}$, they are actually quite small and, as discussed earlier, they do not have any observable consequences.

In addition to the oscillations, the $f^{0}_R = -10^{-1}$ case shows visible differences at late times. These are due to the fact that some of the neglected terms in the perturbed equations are multiplied by time derivatives of $f_R$ and $f_{\rm RR}$, which can become non-negligible for larger values of $|f^{0}_R|$. Such high values of $|f_R^0|$ however, are already ruled by the existing cluster abundance data~\cite{Lombriser:2010mp,Schmidt:2009am} which, combined with other cosmological probes, places a bound of $|f^0_R| \lesssim 10^{-3}$.

The examination of the differences between $\gamma^Q$ and $\gamma^{\rm Ex}$ yields very similar conclusions.  Overall, we see that the quasi-static expressions $\mu^Q$~(\ref{muQ}) and $\gamma^Q$~(\ref{gamQ}) are an excellent representation of $f(R)$ solutions for $|f^{0}_R| \lesssim 10^{-2}$.

 \subsection{$\mu^{\rm BZ}$ and $\gamma^{\rm BZ}$}

 \begin{figure*}[tbph]
\centering
\includegraphics[scale=0.4,trim=0mm 0mm 0mm 0mm]{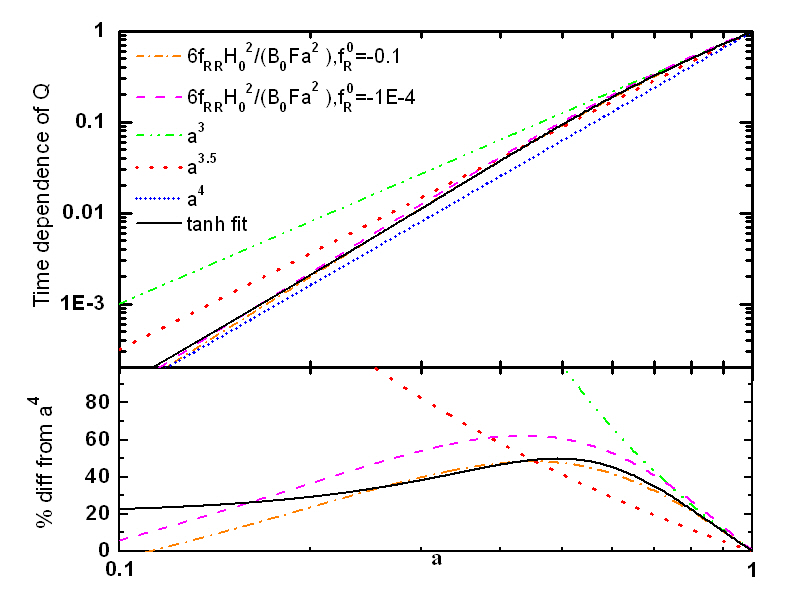}
\caption{Time dependence of $Q$ for $f^{0}_R = -10^{-1}$ (orange dotted-dashed line)  and $f^{0}_R=-10^{-4}$ (purple dashed line), compared to the power laws $a^4$ (blue crossed line), $a^{3.5}$ (red dotted line) and $a^3$ (green double dotted-dashed line) as well as to the phenomenological function $s={[3.5-0.5~ {\rm tanh}(a-0.45)/0.3]}$ (indicated as 'tanh fit',  in solid black). The phenomenological function can follow better the transition from $a^4$ at early times to $a^3$ at late times. Note that $ B_0 \approx - 6 f^{0}_R$.}
\label{fig:mu_plot_log_diff}
\end{figure*}

 \begin{figure*}[tbhp]
\centering
\includegraphics[scale=0.4,trim=0mm 0mm 0mm 0mm]{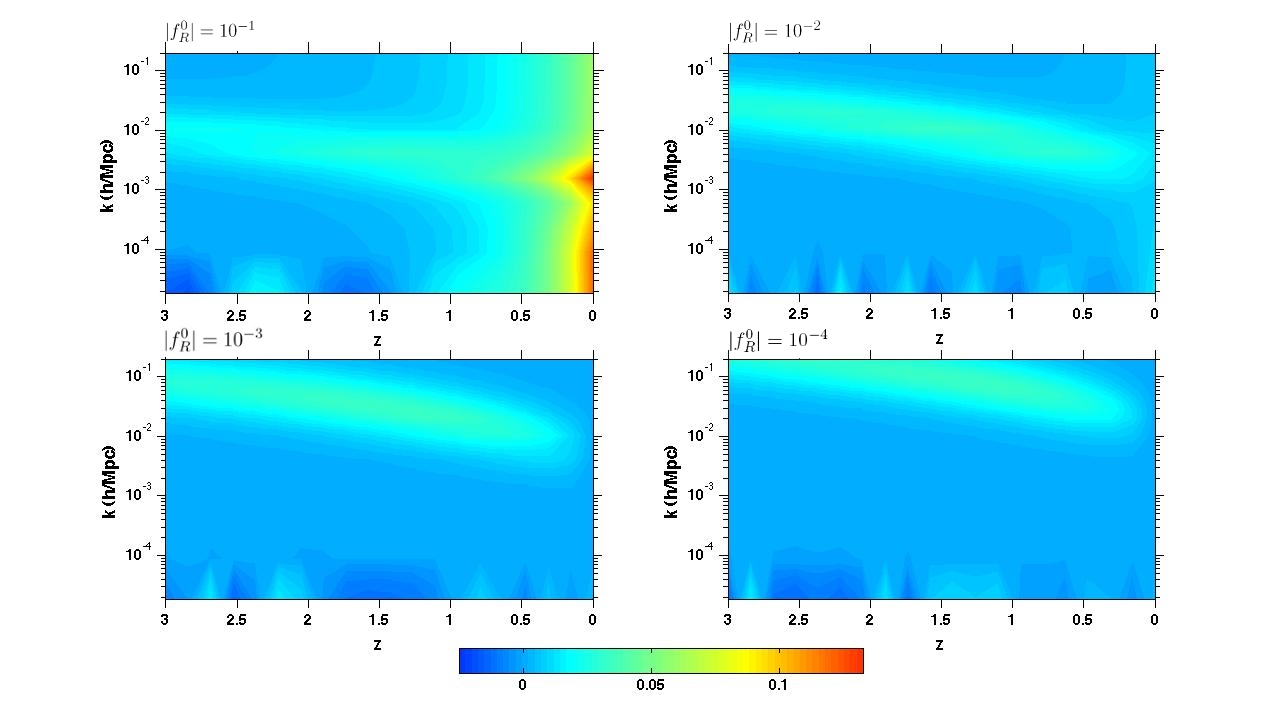}
\caption{The contour plots of the relative difference between $\mu^{\rm BZ}$ with $s=4$ and $\mu^{\rm Ex}$ for four values of $f_R^0$.} 
\label{fig:muBZe}
\end{figure*}

 \begin{figure*}[tbhp]
\centering
\includegraphics[scale=0.4,trim=0mm 0mm 0mm 0mm]{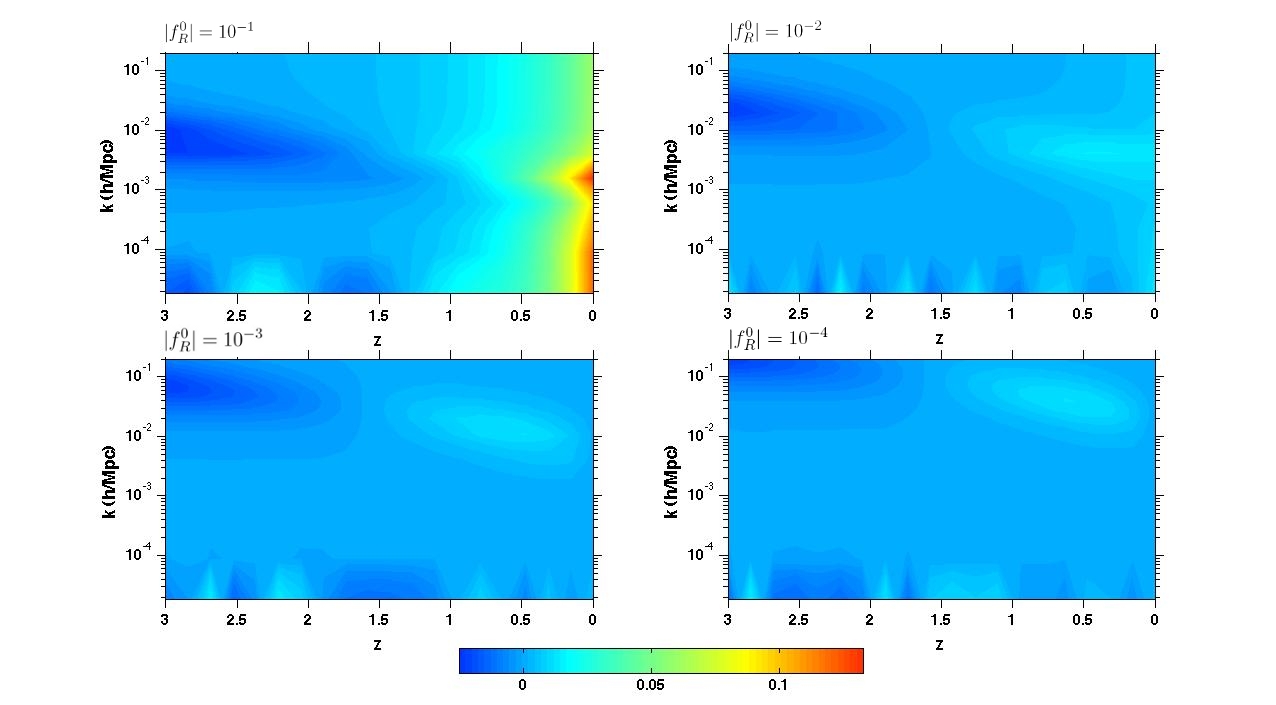}
\caption{The contour plots of the relative difference between $\mu^{\rm BZ}$ with $s=3.5$ and $\mu^{\rm Ex}$ for four values of $f_R^0$.} 
\label{fig:muBZe2}
\end{figure*}

\begin{figure*}[tbhp]
\centering
\includegraphics[scale=0.4,trim=0mm 0mm 0mm 0mm]{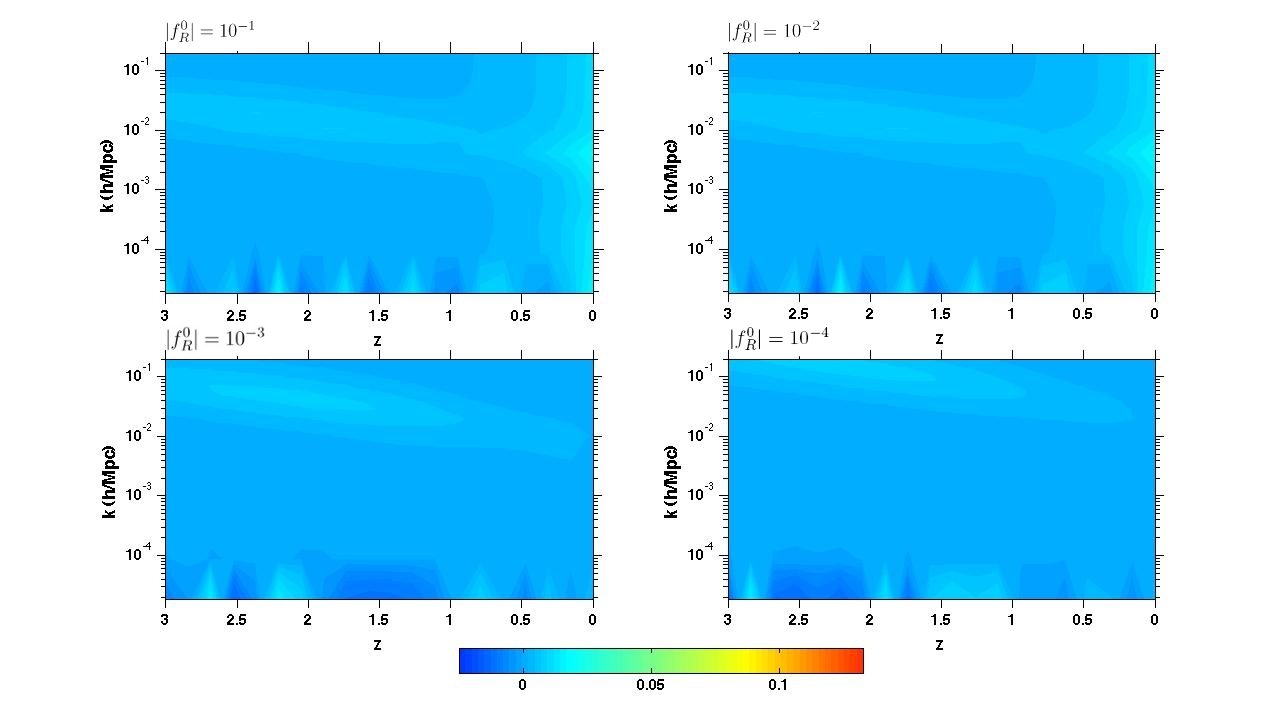}
\caption{The contour plots of the relative difference between $\mu^{\rm BZ}$ with $s=3.5-0.5\tanh[(a-0.45)/0.3]$ and $\mu^{\rm Ex}$ for four values of $f_R^0$.} 
\label{fig:mutanhe_BZ}
\end{figure*}

The BZ parametrization of Eqs.~(\ref{muBZ}) and~(\ref{gamBZ}) is inspired by the quasi-static approximation corresponding to $\mu^Q$ and $\gamma^Q$, but it contains one more degree of approximation -- it assumes a particular form for the time dependence of $Q$. Namely, it sets
\be
{2 \over B_0 {\bar k}^2} Q=\f{6H_0^2}{B_0 }\f{ f_{RR}}{a^2 F} = a^{s} \ .
\label{Q-fit}
\ee
In~\cite{Zhao:2008bn} it was suggested that $s\approx 4$ for a wide range of models. We want to test the accuracy of this assumption and the impact the choice of $s$ makes on $f(R)$ constraints derived using BZ.

\begin{table}[tbph]
\begin{ruledtabular}
\begin{tabular}{c  c c c c c c }	

$f_R^{0}$ & $-10^{-1}$ & $-10^{-2}$ & $-10^{-3}$ & $-10^{-4}$ & $-10^{-5}$ & $-10^{-6}$ \\ 
 & & & & \\
	
$s$ & $3.6$ & $3.6$ & $3.6$ & $3.5$ & $3.3$ & $3.2$  \\
\end{tabular}
\end{ruledtabular}
\caption{The best fit values of $s$, obtained by fitting $a^s$ to the exact time dependence of $Q$ using Eq.~(\ref{Q-fit}), for six values of $f^{0}_R$.}
\label{tb:s_muQ}
\end{table}

We determine numerically the time-dependence of $Q$, and fit $a^s$ to the expression in Eq.~(\ref{Q-fit}). Table~\ref{tb:s_muQ} shows results obtained by fitting over the redshift range $0 \leq z \leq 3$ for models with $10^{-6}\leq|f_R^0|\leq10^{-1}$. We see that $s=4$ is a reasonable choice for $|f^{0}_R| \ge 10^{-3}$, but for smaller $|f^{0}_R|$, values closer to $s=3$ are preferred. Thus, a fixed  $s$, and $s=4$ in particular, does not appear to hold for all $f^{0}_R$. However, while there is a clear dependence of $s$ on $f^{0}_R$, we will argue shortly that such differences should not be taken very seriously.

As already pointed out in~\cite{Thomas:2011}, the time dependence of $Q$ is not exactly a simple power law. This is because the onset of the accelerated expansion at late times changes the evolution of the scalaron mass compared to that in the matter domination epoch. Fig.~\ref{fig:mu_plot_log_diff} shows the numerically reconstructed time dependence of $Q$ for $f^0_R=-10^{-1}$ and $-10^{-4}$, along with plots of $a^s$ with $s=3$, $3.5$ and $4$. Clearly, $s=4$ is a better fit at early times, during the matter dominated epoch, while there is a transition towards $s=3$ near the matter-$\Lambda$ equality. This transition can be modeled using a phenomenological function of the form $s=3.5-0.5\tanh[(a-0.45)/0.3]$, in which $s$ transits from $4$ to $3$ around the time of the matter to $\Lambda$ transition. It turns out, however, that whether one uses $s=4$, $s=3.5$, or the phenomenological function, the differences between $\mu^{\rm Ex}$ and $\mu^{\rm BZ}$ remain roughly of the same order. This can be seen comparing Figs.~\ref{fig:muBZe}, \ref{fig:muBZe2} and \ref{fig:mutanhe_BZ}, where we show the percent difference between $\mu^{\rm BZ}$ with $s=4$ , $s=3.5$ and $s =3.5-0.5\tanh[(a-0.45)/0.3]$ and the numerically found $\mu^{\rm Ex}$. One can see that, while the differences are slightly larger than those between $\mu^Q$ and $\mu^{\rm Ex}$, they follow the same general trend. The most visible difference is due to oscillations in $\mu^{\rm Ex}$ at small $k$. As expected, using the phenomenological functions results in the smallest discrepancies. However, in all three cases, the differences are less than $2\%-3\%$ for $|f^{0}_R| \le 10^{-2}$. The same conclusions hold for $\gamma^{\rm BZ}$.
 
To be more quantitative, we use a Fisher matrix formalism to estimate the variance in $B_0$ given the theoretical error associated with using $\mu^{\rm BZ}$. The Fisher matrix element corresponding to $B_0$ is calculated as
\be
F_{B_0 B_0} = \sum_i \sum_j \frac{\partial \mu^{\rm BZ}_i}{\partial B_0}C_{ij}^{-1}\frac{\partial \mu^{\rm BZ}_j}{\partial B_0} \ ,
\label{mu_fisher}
\ee
where $C_{ij}$ is the covariance matrix for $\mu$, and $\mu^{\rm BZ}_i$ denotes the value of $\mu^{\rm BZ}$ at a particular $(z,k)$ pixel on our grid. We pixelate the $(z,k)$ domain into $20$ bins in $z$ and $20$ bins in $k$, using the same binning configuration as the one used for the PCA analysis in~\cite{Zhao:2009fn,Hojjati:2011xd}. The partial derivatives are calculated analytically using Eq.~(\ref{muBZ}). We take the covariance matrix to be diagonal,
\be
\label{mubz_covmat}
C_{ij} = (\mu_i^{BZ} - \mu_i^{Ex}) \delta_{ij} \ ,
\ee
which effectively treats $\mu_i^{BZ}$ as a biased ``measurement'' of $\mu_i^{Ex}$ that is made independently in each bin. We then take $\sqrt{F^{-1}_{B_0 B_0}}$ as a rough estimate of the variance in $B_0$. 
The variances computed from Eq.~(\ref{mu_fisher}) are two orders of magnitude smaller than their corresponding fducial
values of $B_0$, i.e. 
\be
\sigma_{B_0} \sim 10^{-2} B_0
\ee
for  $10^{-6}<B_0<0.1$.
The forecasted experimental constraints on the $\mu$ (and $\gamma$) from LSST~\cite{Hojjati:2011xd}, based on the same pixelation scheme, are much larger than the systematic theoretical errors due to using $\mu^{\rm BZ}$ and $\gamma^{\rm BZ}$. Hence, we conclude that the BZ approximation can be safely used to put constraints on the $f(R)$ models.\\
We also checked that there is essentially no significant difference  between using $s=4$ versus $s=3.5$, or the phenomenological fit function in the computed values for the variance of $B_0$.
This insensitivity of the uncertainty in $B_0$ to the value of $s$ is because the dependence of $\mu$ and $\gamma$ on $B_0$ is much stronger than its dependence on the time variation. This can be seen by using Eq.~(\ref{muBZ}) to obtain
\begin{equation}
\label{eq:fisher}
\frac{\partial \mu / \partial s}{\partial \mu / \partial B_0} = B_0 \ln a \, ,
\end{equation}
which is small for $B_0 < 0.1$ and $ 0.1 < a < 1$. Similar conclusion hold for $\gamma$. \\
As an additional check, we have implemented $\mu^{\rm Ex}$ and $\gamma^{\rm Ex}$ in MGCAMB~\cite{Hojjati:2011ix} for several representative values of $ 10^{-6} \leqslant |f^{\rm 0}_R| \leqslant 10^{-1}$ and compared the output for the CMB and linear matter power spectra to that obtained using $\mu^{\rm BZ}$. We found that, as expected, the differences are too small to be of relevance.
 
\section{Summary}
\label{sec:summary}

We have examined in some detail the parametrizations commonly employed to evolve linear scalar perturbations in $f(R)$ models and place cosmological constraints. In particular, we have carefully examined the validity of the quasi-static approximation and of the parametrization introduced in~\cite{Bertschinger:2008zb}, (BZ), and adapted to $f(R)$ in~\cite{Giannantonio:2009gi}. 
 
After reviewing the main features of the linearized Einstein equations in $f(R)$, we have analyzed near horizon effects and provided an analytical insight on the oscillations that are observed at early times on large scales. Upon showing that the oscillations originate from the scalaron, and linking their amplitude and frequency to the mass of the latter, we have argued that they have no observable signatures. We have then numerically analyzed the accuracy of the quasi-static approximation, finding that on linear sub-horizon scales the functions $\mu^Q$ and $\gamma^Q$ describe very closely the modified Poisson and anisotropy equations. \\
Furthermore, we have analyzed the BZ parametrization~\cite{Bertschinger:2008zb}, in the form adapted to $f(R)$, and in particular the validity of its approximation of the time-dependence of $Q$ as a power law $a^s$. We have found that there is not a fixed value of $s$ valid for all designer $f(R)$ models, but rather the best fit value of $s$ depends on $f_R^0$ and the range of redshifts considered. A comparison with the exact time dependence of $Q$ shows that $s \approx 4$ is a good fit for early times, while $s\approx 3$ fits better at late times. The transition of the value of $s$ can be modeled using a phenomenological function that takes into account the transition from matter to $\Lambda$ dominated era. However, we have shown that $\mu$ and $\gamma$ are relatively insensitive to the variation of $s$ so that $s=3.5$, $s=4$, or the phenomenological fit lead to a theoretical systematic error in $\mu$ (and $\gamma$) that is much smaller than the observational uncertainty from a future survey like LSST. Hence, the BZ parametrization can be safely used for deriving constraints on $f(R)$ models from upcoming large scale surveys. 

\acknowledgements We benefited from prior collaborations and discussions with Edmund Bertschinger, Kazuya Koyama, and Gong-Bo Zhao, and from discussions with Tessa Baker, Pedro Ferreira, Andrei Frolov and Constantinos Skordis. AH is supported by an NSERC Discovery Grant and partly by World Class University grant R32-2009-000-10130-0 through the National Research Foundation, Ministry of Education, Science and Technology of Korea, LP and ST by an NSERC Discovery grant and funds from SFU, and AS by the grant NSF AST-0708501 and the SISSA Excellence Grant.

\end{document}